\documentclass[12pt]{iopart}
\usepackage{graphicx}
\usepackage{cite}
\usepackage{amsfonts}
\begin{document}

\title[On the predictive power of Local Scale Invariance]
      {On the predictive power of Local Scale Invariance}
\author{Haye Hinrichsen}
\address{Universit\"at W\"urzburg\\
	 Fakult\"at f\"ur Physik und Astronomie\\
         D-97074 W\"urzburg, Germany}
\ead{hinrichsen@physik.uni-wuerzburg.de}

\def\d{\mathrm{d}}

\begin{abstract}
Local Scale Invariance (LSI) is a theory for anisotropic critical phenomena designed in the spirit of conformal invariance. For a given representation of its generators it makes non-trivial predictions about the form of universal scaling functions. In the past decade several representations have been identified and the corresponding predictions were confirmed for various anisotropic critical systems. Such tests are usually based on a comparison of two-point quantities such as autocorrelation and response functions. The present work highlights a potential problem of the theory in the sense that it may predict \textit{any} type of two-point function. More specifically, it is argued that for a given two-point correlator it is possible to construct a representation of the generators which exactly reproduces this particular correlator. This observation calls for a critical examination of the predictive content of the theory.
\end{abstract}

\submitto{Journal of Statistical Mechanics: Theory and Experiment}


\parskip 2mm 

\section{Introduction}

\textit{Local scale invariance} (LSI) stands for a theory developed by M. Henkel and collaborators which generalizes global scale invariance of anisotropic critical systems and ageing phenomena to a local space-time-dependent symmetry~\cite{Henkel02a}. It is inspired by the success of conformal invariance applied to two-dimensional equilibrium systems~\cite{Polyakov70a,Henkel99} and uses a similar terminology. 

Local scale transformations for anisotropic systems are generated by an infinite set of generators $X_{-1},X_0,X_1,X_2,\ldots$ and $Y_{-1/z},Y_{-1/z+1},Y_{-1/z+2},\ldots$. These generators obey the commutation relations
\begin{eqnarray}
\label{XXComm} 
[X_n,X_m]&=&(n-m)X_{n+m}\\[0mm]
\label{XYComm}
[X_n,Y_m]&=&\Bigl(\frac{n}{z}-m\Bigr) Y_{n+m}\,,
\end{eqnarray}
where $z$ is the dynamical exponent which quantifies the degree of anisotropy. 

Remarkably, the theory predicts the specific form of universal scaling functions appearing in response or correlation functions while it makes no prediction about the values of critical exponents. In recent years the theory has been extended and successfully applied to a large variety of models~\cite{HenkelPleimling03a,PiconeHenkel04a,HenkelEtAl04a,RamascoEtAl04a,EnssEtAl04a,HenkelPleimling05a,BaumannEtAl05a,StoimenovHenkel05a,BaumannEtAl06a,HenkelUnterberger06a,Odor06a,HenkelEtAl06a,HenkelPleimling06a,RothleinEtAl06a,BaumannHenkel07a,LorenzJanke07a,BaumannEtAl07b,Henkel07a}. This success raised the hope that LSI could be a \textit{generic} symmetry of scale-invariant anisotropic systems. However, some authors reported results which seem to be incompatible with the predictions of LSI~\cite{CalabreseGambassi03a,CorberiEtAl03a,PleimlingGambassi05a,LippielloEtAl06,Hinrichsen06,BaumannGambassi07a,PaulEtAl07,Hinrichsen08a} which released a debate concerning the applicability of the theory. 

In a given model local scale invariance can be established by choosing a suitable representation of the algebra~(\ref{XXComm})-(\ref{XYComm}) in such a way that the system under consideration is invariant under transformations generated by $X_n$ and $Y_m$. In some cases such a representation can be derived exactly from an underlying partial differential equation, whereas in other cases LSI is just assumed as a hypothetical symmetry, leading to certain predictions which can be tested by numerical simulations. Since LSI is a model-independent theory these predictions depend exclusively on the chosen representation of its generators. 

The simplest representation can be derived from the Schr{\"o}dinger equation which describes diffusing particles with $z=2$~\cite{Henkel02a}. This representation was generalized to the case $z\neq 2$ and successfully testet for $z=4$~\cite{Henkel07a}. Recently, Baumann and Henkel~\cite{HenkelBaumann07b} found two representations of the LSI algebra~(\ref{XXComm})-(\ref{XYComm}) for arbitrary $z$ which involve non-local fractional derivatives~\cite{Gelfand64a} (see Appendix B). The present work generalizes this concept even further by considering representations with \textit{arbitrary} non-local operators. It is shown that the resulting representations of the LSI algebra are so general that in principle any two-point correlation function can be reproduced by the theory. This has important consequences regarding the predictive power of the theory, as will be discussed at the end of this paper.

\section{Geometrical interpretation of the generators}

To gain some intuition how the generators $X_n$ and $Y_m$ work let us first recall the geometrical interpretation of the generated transformations. For simplicity, we will restrict to the 1+1-dimensional case. The generalization to higher dimensions is not difficult and requires to replace $Y_m$ by a vector operator $Y_m^{(j)}$, see Refs.~\cite{Henkel02a,Roger06a}.

\begin{figure}[t]
\includegraphics[width=150mm]{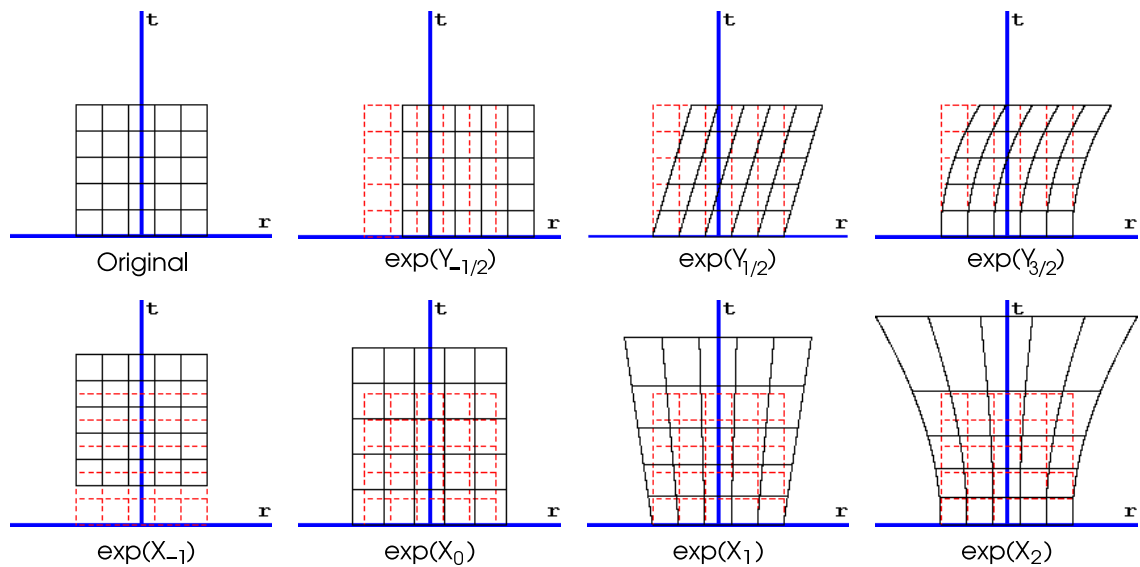}
\caption{Distortion of a square lattice in 1+1 dimensions through local scale transformations generated by various LSI generators, here shown for the case $z=2$. Qualitatively similar deformations are observed for any $z>1$.}
\label{fig:lsi}
\end{figure}

The simplest representation of the generators $X_n$ and $Y_m$, which describes the geometrical content of local scale transformations, is given by
\begin{eqnarray} 
\label{Xraw}
X_n &=& -t^{n+1} \partial_t - \frac{n+1}{z}t^{n}r\partial_r\,,\\
\label{Yraw}
Y_{k-1/z} &=& -t^k \partial_r\,,
\end{eqnarray}
where we have adopted the convention of non-integral indices $m=k-1/z$ with $k\in \mathbb{N}$. One can easily verify that these operators satisfy the commutation relations~(\ref{XXComm}) and~(\ref{XYComm}). Moreover, one can see that they carry the physical dimensions
\begin{eqnarray} 
[X_n] &=& [\mbox{time}]^n\\[0mm]
[Y_{k-1/z}] &=& [\mbox{time}]^k [\mbox{length}]^{-1}\,.
\end{eqnarray}
As shown in Fig.~\ref{fig:lsi}, each of these generators corresponds to a well-defined geometrical transformation in space-time. For example, the generator $X_{-1}=-\partial_t$ generates translations in time while $Y_{-1/z}=-\partial_r$ generates translations in space:
\begin{eqnarray}
\exp(\tau X_{-1})\,f(t,r)&=&f(t-\tau,r)\,,\\
\exp(s Y_{-1/z})\,f(t,r)&=&f(t,r-s).
\end{eqnarray}
%
%
\noindent
Likewise $X_0=-t\partial_t-\frac{1}{z}r\partial_r$ generates anisotropic dilatations 
\begin{equation}
\label{bareX0}
\exp(\lambda X_0)f(t,r)=f(t/b,\,r/b^{1/z})
\end{equation}
by the factor $b=e^\lambda$ and thus it can be identified as the generator of \textit{global} scale trans\-formations. Here the amount of anisotropy is controlled by the dynamical exponent~$z$. 

The lowest non-trivial operators, which mix space and time, are $X_1$ and $Y_{1-1/z}$. As shown in Fig.~\ref{fig:lsi}, the operator $Y_{1-1/z}$ generates a Galilei transformation
\begin{equation}
\exp(c Y_{1-1/z})f(t,r)=f(t,\,r-ct)\,
\end{equation}
which may be interpreted as a global \textit{shear transformation} in space-time. Similarly, $X_1$ generates a dilatation with an elongation factor proportional to the actual time. By combining the action of all generators it is possible to generate the full group of local scale transformations. Note that these transformations obey causality, i.e., in Fig.~\ref{fig:lsi} horiztonal lines will always be mapped onto horizontal lines.

\section{Local scale invariance of quasi-primary fields}

\begin{figure}[t]
\includegraphics[width=155mm]{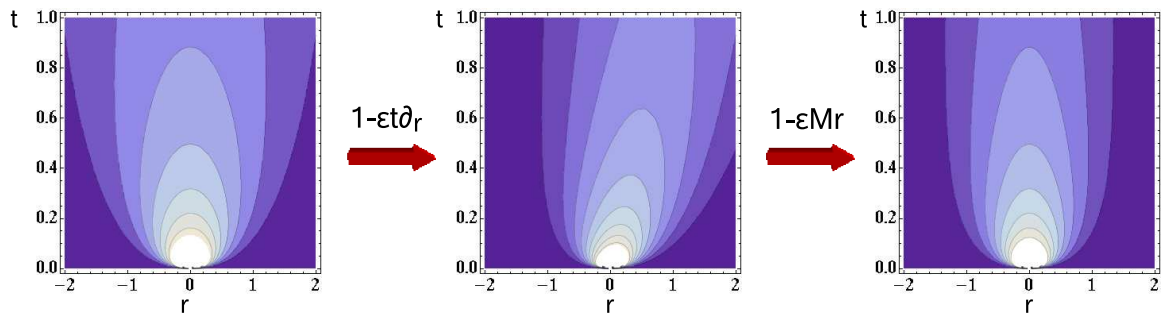}
\caption{Action of the generator $Y_{-1/2}$ using the example of a simple diffusion equation $\partial_t \phi(r,t) = D \nabla^2 \phi(r,t)$. In this case the two-point response function is an ordinary Gaussian distribution $G(r,t)=\frac{1}{2\sqrt{\pi D t}} \exp(-\frac{r^2}{4 D t})$ which is shown in the left panel. As can be seen, this function is \textit{not} invariant under infinitesimal transformations generated by $Y_{1/2}=-t\partial_r$ since geometrical shear leads to a skewed function of the form $G(r,t)(1+\frac{r\epsilon}{2D})$. Therefore, the generator $Y_{1/2}=-t\partial_r$ is not a symmetry operator, rather it has to be extended by an additional term that compensates this tilt, restoring the original form of the function. For the diffusion equation this term takes the simple form $-\mathcal Mr$ with the so-called mass $\mathcal M=\frac{1}{2D}$. As shown in the right panel, a subsequent application of this term restores the original non-tilted function up to order $\mathcal O(\epsilon^2)$. This demonstrates the mechanism which makes the diffusion equation invariant under the action of the generator $Y_{1/2}=-t\partial_r-\mathcal M r$.}
\label{fig:schroedinger}
\end{figure}

As a next step, the geometric generators~(\ref{Xraw})-(\ref{Yraw}) have to be extended in order to describe the symmetry properties of a scale-free critical phenomenon. Within the framework of LSI it is assumed that the physical properties of such a system can be expressed in terms of so-called \textit{quasi-primary fields}\footnote{Presently it is not yet fully clear how quasi-primary fields can be characterized in anisotropic systems. Usually it is believed that order parameter fields are quasi-primary while their derivatives are not. For a discussion see Ref.~\cite{HenkelEtAl06a}.}, here denoted as $\phi(t,r)$, which transform covariantly under the action of the generators. As we will see, this requires to extend the representation of the generators by additional terms.

For simplicity let us assume that the system under consideration is translational invariant in space and time, for example a critical kinetic Ising model in its stationary equilibrium state. This means that all quasi-primary fields are translational invariant as well, hence the two operators $X_{-1}=-\partial_t$ and $Y_{-1/z}=-\partial_r$ are already symmetry operators and need no extension. The situation is different for global dilatations generated by $X_0$. Here a quasi-primary field changes its amplitude according to the scaling law
\begin{equation}
\phi\bigl(t/b,r/b^{1/z}\bigr) = b^{x/z} \phi(t,r)\,,
\end{equation}
where the exponent $x$ is the so-called \textit{scaling index} associated with the field $\phi(t,r)$. Consequently, the bare generator $X_0$ defined in~(\ref{bareX0}) is no longer a symmetry operator, rather it has to be extended by a suitable term that compensates the change in the field amplitude. Obviously the required term is just a constant, leading to the standard representation
\begin{equation}
\label{X0rep}
X_0 \;=\, -t\partial_t -\frac{1}{z} r\partial_r -\frac{x}{z}\,.
\end{equation}
Next, let us consider the operator $Y_{1-1/z}$ which generates global shear transformations. It is important to note that global shear \textit{itself} is generally no symmetry transformation because it distorts the fields in a non-trivial way. This is demonstrated in Fig.~\ref{fig:schroedinger} using the example of the diffusion equation where $z=2$. As can be seen, global shear distorts the response function, leading to a skewed profile in space-time. Therefore, the generator $Y_{1-1/z}$ has to be extended by a suitable additional term that compensates this distortion. For the diffusion equation this term takes the particularly simple form $-\mathcal M r$, where $\mathcal M$ is the so-called mass parameter. However, in general the required compensation terms may be much more complicated.

To be as general as possible, we therefore assume that $Y_{1-1/z}$ is extended by an arbitrary linear operator $\mathcal{B}_{1-1/z}$:
\begin{equation}
Y_{1-1/z} \;=\; -t \partial_r - \mathcal{B}_{1-1/z}\,.
\end{equation}
Likewise we assume that all remaining generators, which generate combinations of shear and dilatations, are extended by certain linear operators as well:
\begin{eqnarray} 
X_n &=& -t^{n+1} \partial_t - \frac{n+1}{z}t^{n}r\partial_r-\frac{x}{z}(n+1)t^n - \mathcal{A}_n
\hspace{5mm} n=1,2,3,\ldots\\
Y_{k-1/z} &=& -t^k \partial_r - \mathcal{B}_{k-1/z}
\hspace{52mm} k=1,2,3,\ldots
\end{eqnarray}
The linear operators $\mathcal{A}_n$ and $\mathcal{B}_m$, by which the generators are extended, are of course not independent, rather they are constrained by the commutation relations. For example, the operator $\mathcal A_2$ cannot be chosen freely, instead it is constrained by the commutation relations
\begin{eqnarray}
\label{constraints}
&[X_2,X_0]=2X_2 &\quad\Longrightarrow\quad [\mathcal A_2,X_0]=2\mathcal A_2\\
&[[X_2,X_{-1}],X_{-1}]=6X_0&\quad\Longrightarrow\quad [[\mathcal A_2,X_{-1}],X_{-1}]=0\,.
\end{eqnarray}
%

\section{Iterative construction of the generators}

The first point of this work is to show that any space-time representation of the LSI commutation relations is fully determined by the generator $X_2$ or, equivalently, by the operator $\mathcal A_2$.\footnote{If one is only interested in the subalgebra $\{X_{-1},X_0,X_1,Y_{-1/z},Y_{1-1/z}\}$ it even suffices to specify $\mathcal A_1$.}

The construction starts by choosing the operator $\mathcal A_2$ in such a way that it obeys the constraints~(\ref{constraints}). Once $\mathcal A_2$ is specified, all other operators can be constructed iteratively as follows. The lowest generators $X_{-1}$, $X_0$ and $Y_{-1/z}$ are always given by their standard representation
\begin{eqnarray}
X_{-1} 	&:=& -\partial_t \\
X_0 	&:=& -t\partial_t -\frac{1}{z} r\partial_r -\frac{x}{z} \\
Y_{-1/z} &:=& = -\partial_r\,.
\end{eqnarray}
Moreover, the generator $X_1$ can be computed by setting
\begin{equation}
\label{ConstructX1}
X_1 := \frac13\,[X_2,X_{-1}]
\end{equation}
meaning that $\mathcal A_1=\frac13[\partial_t,\mathcal A_2]$. Likewise, all other generators can be constructed recursively by
\begin{eqnarray}
\label{ConstructX}
X_n &:=& \frac{1}{n-2}\,[X_{n-1},X_1]\,\hspace{17mm} n=3,4,5,\ldots\\
\label{ConstructY}
Y_m &:=& \frac{1}{m-1/z-1}\,[Y_{m-1},X_1]\,\hspace{4mm} m=1-1/z,\,2-1/z,\,3-1/z,\ldots
\end{eqnarray}
As shown in Appendix A, any set of generators, which is contructed in such a way, satisfies all commutation relations in Eqs.~(\ref{XXComm}) and~(\ref{XYComm}) automatically. Therefore, we can conclude that any representation is fully determined by a single linear operator, namely,~$\mathcal A_2$.

\section{Representation by integral kernels}

Given that the whole representation is determined by the linear operator $\mathcal A_2$, one has to specify the most general form of this operator under the constraints~(\ref{constraints}). To this end it is convenient to represent $\mathcal A_n$ acting on some function $\phi$ in form of a convolution integral 
\begin{equation}
[\mathcal A_n\phi](t,r) \;=\;
\int \d t' \int \d r'\, A_n(t,r,t',r')\phi(t',r')
\end{equation}
with a kernel $A_n(t,r,t',r')$ which can be thought of as the `matrix elements' of $\mathcal A_n$. Likewise, the operators $\mathcal B_m$, which appear in the generator $Y_m$, can be written as convolution integrals
\begin{equation}
[\mathcal B_m\phi](t,r) \;=\;
\int \d t' \int \d r'\, B_m(t,r,t',r')\phi(t',r')\,
\end{equation}
with a kernel $B_m(t,r,t',r')$. The local contributions of these kernels, which appear as ordinary differential operators in the LSI representation, correspond to Dirac $\delta$-functions and their derivatives. However, in general the kernel may be non-local in space and time, including the recently discovered non-local representations involving fractional derivatives as special cases. 

Because of the commutation relation $[X_n,X_0]=nX_n$ the kernel itself is a generalized homogeneous function under anisotropic dilatation by a factor $b>0$:
\begin{equation}
\label{HomA}
A_n(t/b,\, r/b^{1/z},\, t'/b,\, r'/b^{1/z}) \;=\; b^{1/z+1-n} A_n(t,r,t',r')\,.
\end{equation}
Moreover, the kernel $A_2$ is constrained by $[[X_2,X_{-1}],X_{-1}]=0$ in Eq.~(\ref{constraints}), tantamount to $[[\mathcal A_2,\partial_t],\partial_t]=0$, which implies that $A_2$ has to be of the form
\begin{equation}
\label{A2}
A_2(t,r,t',r') \;=\; \frac32(t+t')K(t-t',r,r') + L(t-t',r,r'),
\end{equation}
where $K$ and $L$ are functions which depend only on the time difference $t-t'$. Because of Eq.~(\ref{HomA}), they are generalized homogeneous functions as well:
\begin{eqnarray}
\label{ScalingK}
K\Bigl(\frac{t-t'}{b},\, \frac{r}{b^{1/z}},\, \frac{r'}{b^{1/z}}\Bigr) &=& b^{1/z}  \, K(t-t',r,r')\,,\\
L\Bigl(\frac{t-t'}{b},\, \frac{r}{b^{1/z}},\, \frac{r'}{b^{1/z}}\Bigr) &=& b^{1/z-1}\, L(t-t',r,r')\,.
\end{eqnarray}
Hence any representation of the LSI algebra is determined by two homogeneous time-translation-invariant functions $K$ and $L$.\\

Because of $X_1=\frac13[X_2,X_{-1}]$ the kernel $\mathcal A_1=-\frac13[\mathcal A_2,\partial_t]$ is essentially the temporal derivative of $A_2$, i.e.
\begin{equation}
A_1(t,r,t',r') \;=\; K(t-t',r,r')\,.
\end{equation}
Because of Eq.~(\ref{XYComm}), the operators $\mathcal B_m$ are given by
\begin{equation}
\mathcal B_{k-1/z}=\frac{z}{k+1}[\mathcal A_k,\partial_r]\,,\qquad\qquad k=0,1,2,\ldots
\end{equation}
so that the kernel of the shear generator $Y_{1-1/z}$ reads
\begin{equation}
B_{1-1/z}(t,r,t',r') = \frac{z}{2}\Bigl(\frac{\partial}{\partial_r}+\frac{\partial}{\partial_{r'}}\Bigr)K(t-t',r,r')\,.
\end{equation}
Again this kernel is a generalized homogeneous function, i.e.,
\begin{equation}
B_{1-1/z}\Bigl(\frac{t}{b},\, \frac{r}{b^{1/z}},\, \frac{t'}{b},\, \frac{r'}{b^{1/z}}\Bigr) \;=\;
 b^{2/z}  \, B_{1-1/z}(t,r,t',r')\,.
\end{equation}
Once all kernels have been determined, one can easily check them by testing the commutation relations $[X_n,X_{-1}]=(n+1)X_{n-1}$ and $[Y_n,X_{-1}]=(n+1/z)Y_{n-1}$, which can be translated into the differential equations
\begin{eqnarray}
\Bigl(\frac{\partial}{\partial_t}+\frac{\partial}{\partial_{t'}}\Bigr) A_n(t,r,t',r') &=& (n+1) A_{n-1}(t,r,t',r')\,, \\
\Bigl(\frac{\partial}{\partial_t}+\frac{\partial}{\partial_{t'}}\Bigr) B_n(t,r,t',r') &=& (n+1/z) B_{n-1}(t,r,t',r')\,.
\end{eqnarray}
%
\section{Two-point correlation functions}

Let us now investigate the properties of a correlation function
\begin{equation}
\label{corr}
C(t,r,\,t',r') \;=\; \langle \phi_1(t,r) \phi_2(t',r') \rangle\,\qquad \qquad (t>t')
\end{equation}
of two quasi-primary fields $\phi_1$ and $\phi_2$. For simplicity let us assume that both fields carry the same scaling dimension $x_1=x_2=x$ and that the correlator respects causality, i.e., it is nonzero only for $t\geq t'$. Because of translational invariance in space and time this function depends only on the differences of the coordinates. Moreover, invariance under global scale transformations, as expressed by the condition $X_0C=0$, implies the scaling form
\begin{equation}
\label{ScalingFormC}
C(t,r,\,t',r') \;=\; \Theta(t-t')\,(t-t')^{-2x/z}\,\Phi\Bigl(\frac{r-r'}{(t-t')^{1/z}}\Bigr)\,,
\end{equation}
where $\Phi$ is a scaling function and $\Theta(t-t')$ is the Heaviside step function which accounts for causality. 

In the following we first address the problem how $\Phi$ can be determined for a given representation of LSI generators. Then we consider the inverse problem, i.e., for a given correlation function we ask for a suitable representation of the generators.

\subsection{Integro-differential equation for the scaling function}

LSI is based on the postulate that correlation functions of primary fields are invariant under the action of the generators, i.e., $X_nC=Y_mC=0$. These conditions lead to integro-differential equations which determine the form of the scaling function $\Phi$. It is important to note that in the LSI theory the representation of the generators acting on $\phi_1$ and $\phi_2$ are generally different. More specifically, it was argued that a correlator vanishes unless the two representations are related in a specific way, giving rise to so-called Bargmann superselection rules. For example, in the standard Schr{\"o}dinger representation the `mass terms' occuring in the generators are known to have different signs. For this reason we will work with two different representations, denoting the integral kernels acting on $\phi_1$ and $\phi_2$ by the superscripts $^{(1)}$ and $^{(2)}$, respectively.

Let us first consider the generator $X_1$, which acts on the two-point function by
\begin{eqnarray}
\label{convint}
&0&=[X_1C](t,r,\,t',r')\nonumber \\ &&=-\Bigl(t^2\partial_t+{t'}^2\partial_{t'}+\frac{2tr}{z}\partial_r+\frac{2t'r'}{z}\partial_{r'}+\frac{2xt}{z}+\frac{2xt'}{z}\Bigr) \, C(t,r,t',r')\\ \nonumber
&&\hspace{5mm}- \int_{t'}^t \d t'' \int_{-\infty}^\infty dr'' K^{(1)}(t-t'',r,r'') \, C(t'',r'',t',r')\\ \nonumber
&&\hspace{5mm}- \int_{t'}^t \d t'' \int_{-\infty}^\infty dr'' K^{(2)}(t'-t'',r',r'') \, C(t,r,t'',r'')\nonumber\,.
\end{eqnarray}
Inserting the scaling form~(\ref{ScalingFormC}) and using the homogeneity condition~(\ref{ScalingK}) one obtains an integro-differential equation. As we have assumed the correlator to be translational invariant, this equation is generally over-determined unless the kernels $K^{(1,2)}$ obey specific constraints, referred to as Bargmann superselection rules. A straight-forward calculation shows that one obtains an autonomous integro-differential equation for the scaling function $\Phi$ if and only if the two kernels have the form
\begin{eqnarray}
\label{K12}
K^{(1)}(t-t',r,r') &=& \hspace{2mm}\frac14(r+r')^2\,f(t-t',r-r')\\
&&+\frac12(r+r')\,g(t-t',r-r')+h(t-t',r-r')\nonumber\\
K^{(2)}(t-t',r,r') &=& \hspace{2mm}-\frac14(r+r')^2\,f(t'-t,r'-r)\\
&&+\frac12(r+r')\,g(t'-t,r'-r)-h(t'-t,r'-r)\nonumber\,,
\end{eqnarray}
where $f,g,h$ are certain functions which depend on only two parameters. Because of Eq.~(\ref{ScalingK}) these functions are homogeneous, hence they obey the scaling form.
\begin{eqnarray}
\label{fgscaling}
f(\tau,\xi)&=&\tau^{-3/z}F(\xi\tau^{-1/z})\,,\nonumber \\
g(\tau,\xi)&=&\tau^{-2/z}G(\xi\tau^{-1/z})\,,\\
h(\tau,\xi)&=&\tau^{-1/z}H(\xi\tau^{-1/z})\,.\nonumber
\end{eqnarray}
With these kernels and the scaling form~(\ref{ScalingFormC}) the integral equation~(\ref{convint}) turns into an integro-differential equation for the scaling function $\Phi(\xi)$:
\begin{eqnarray}
\label{IDGL}
\Phi '(\xi) &\,+ \,&
z \int_{-\infty}^\infty \d{\tilde\xi} \int_0^1 \d\mu \,\mu ^{\frac{1-2 x}{z}}
   \Bigl[{\tilde\xi}  (1-\mu)^{-3/z} \mu
   ^{1/z}F \Bigl(\frac{\xi -\mu
   ^{1/z} {\tilde\xi} }{(1-\mu)^{1/z}}\Bigr) 
+\\&&  \hspace{43mm} (1-\mu)^{-2/z} G\Bigl(\frac{\xi -\mu
   ^{1/z} {\tilde\xi} }{(1-\mu)^{1/z}}\Bigr)\Bigr]\;
   \Phi ({\tilde\xi} ) \;=\; \nonumber 0\,.
\end{eqnarray}
As can be seen, the kernel function $H$ drops out, meaning that it does not influence the form of two-point functions. Knowing the kernels $F,G$, this is the integro-differential equation which determines the scaling function $\Phi$ of the two-point correlation function.

Turning to the generator of shear transformations $Y_{1-1/z}$, one can show that the invariance condition $[Y_{1-1/z}C](t,r,\,t',r')=0$ leads exactly to the same integral equation and thus does not provide any new information.

\subsection{Inverse problem}

Let us now consider the inverse problem which can be formulated as follows: For a given two-point function characterized by $x$, $z$, and $\Phi$ we would like to determine appropriate kernel functions $K^{(1,2)}$ such that $\Phi$ is a solution of Eq.~(\ref{IDGL}). Since these kernel functions determine all LSI generators through recursion relations, solving the inverse problem would mean to construct a representation that renders exactly a given two-point function. In the following we outline how this problem may be solved.

For simplicity let us restrict to the case $G=H=0$, i.e. we want to determine $F$ for a given $\Phi$. With the substitution $\tilde\xi\to\mu^{-1/z}\tilde\xi$ the integro-differential equation~(\ref{IDGL}) turns into a convolution product
\begin{equation*}
\Phi '(\xi) +
z \int_{-\infty}^\infty \d{\tilde\xi} \int_0^1 \d\mu \,\mu ^{-2x/z}
    (1-\mu)^{-3/z} \,{\tilde\xi}\, F \Bigl(\frac{\xi -{\tilde\xi} }{(1-\mu)^{1/z}}\Bigr) 
   \Phi ({\mu
   ^{-1/z} \tilde\xi} ) \;=\; \nonumber 0\,.
\end{equation*}
By introducing the Fourier transforms
\begin{eqnarray}
&&F (\alpha) \label{FTF}
\;=\; \frac{1}{\sqrt{2\pi}}\int_{-\infty}^{+\infty}\d k\, e^{i k \alpha}\,
\tilde F(k) \,,\\ \label{FTPHI}
&&\Phi(\xi) \;=\;\frac{1}{\sqrt{2\pi}}\int_{-\infty}^{+\infty}\d k\,e^{i k \xi}\,\tilde \Phi(k) 
\end{eqnarray}
one is led to 
\begin{eqnarray}
&& \frac{1}{\sqrt{2\pi}} \int_{-\infty}^{+\infty}\d k\, \,e^{i k \xi} \,ik\,\tilde\Phi(k) \;-\;\\&&
\frac{z}{2\pi} \int_0^1\d\mu\,\mu^{\frac{1-2x}{z}}(1-\mu)^{-2/z}\nonumber
\int_{-\infty}^{+\infty}\d\tilde\xi \int_{-\infty}^{+\infty}\d k_1 \int_{-\infty}^{+\infty}\d k_2\,
\tilde\Phi(\mu^{1/z}k_2)\times \\&&\hspace{50mm}\times 
\tilde F\biggl((1-\mu)^{1/z}k_1\biggr)\,\frac{1}{i}\frac{\partial}{\partial{k_2}}\,
e^{ik_1(\xi-\tilde\xi)+i k_2 \tilde \xi}\;=\;0\,.\nonumber
\end{eqnarray}
After integrating by parts in $k_2$ one can carry out the integration over $\tilde \xi$. Finally another Fourier transformation of the entire equation yields
\begin{equation}
\frac{k \tilde\Phi(k) }{\sqrt{2\pi}} \;+\;
z \int_0^1\d\mu\,\mu^{\frac{2-2x}{z}}(1-\mu)^{-2/z}\tilde F\biggl((1-\mu)^{1/z}k\biggr)\, \tilde\Phi'(\mu^{1/z}k)
\;=\;0\,.
\end{equation}
This equation holds for all $k\in \mathbb{R}$. As $\Phi$ and $F$ are symmetric, let us restrict to $k>0$. Substituting $\lambda=\mu^{1/z}k$ one obtains an inhomogeneous Volterra integral equation of the first kind for $\tilde F(\lambda)$
\begin{equation}
\label{Volterra}
\tilde\Phi(k)\;=\; 
\int_0^k \d\lambda\,\Psi(k,\lambda)\,\tilde F(\lambda)
\end{equation}
with the kernel
\begin{equation}
\label{VKernel}
\Psi(k,\lambda) \;=\; \sqrt{2\pi} \, z^2\, k^{1-z}
\,\lambda^{3-z} \, \Bigl[1-\lambda^zk^{-z}\Bigr]^{\frac{2-2x}{z}} \,
\tilde\Phi' \Bigl(\Bigl[1-\lambda^zk^{-z}\Bigr]^{1/z} k\Bigr)\,.
\end{equation}
Hence for given $x,z$, and $\Phi$ the solution of the inverse problem amounts to perform the following non-trivial steps:
\begin{enumerate}
\item Compute the Fourier transform of $\Phi$ in Eq.~(\ref{FTPHI}).
\item Plug $\tilde \Phi$ into Eq.~(\ref{VKernel}), compute the kernel $\Psi(k,\lambda)$, and solve the integral equation~(\ref{Volterra}) to obtain $\tilde F$.
\item Compute the inverse Fourier transform of $\tilde F$, plug it into Eqs.~(\ref{fgscaling}) and (\ref{K12}) and determine the kernel $K$. This kernel establishes the representation of $X_1$.
\item Determine the LSI generators by recursion, setting e.g. $L=0$.
\end{enumerate}
%

\section{Discussion}

The success of LSI applied to various exactly solvable system raised the hope that this theory might be a generic feature of anisotropic scale-free phenomena, including systems which cannot be solved exactly. To verify this expectation, various authors measured two-point autocorrelation and response functions numerically and compared them with the predictions of LSI using a suitable representation of its generators. For some systems, most notably the kinetic Ising model and the contact process, deviations were found, leading the authors to the conclusion that those systems are probably not invariant under local scale transformations.

However, such a conclusion would be premature. The results of the preceding section lead to the conjecture that for any physically meaningful two-point function with arbitrary $x$ and $z$ it is possible to construct a representation of LSI generators which precisely reproduces the desired two-point function. On the one hand, this means that numerical discrepancies do not necessarily falsify LSI, instead they could also come from choosing the wrong representation. On the other hand, the space of possible representations is so huge that LSI itself can probably not predict the form of two-point functions, hence on this level it cannot be used to set up a classification scheme of anisotropic critical phenomena.

For the sake of simplicity the present work was restricted to stationary situations in 1+1 dimensions, where correlation functions depend only on differences of the coordinates. It would be interesting to investigate recent applications of LSI to ageing~\cite{Struik78a,Young98,ZippoldEtAl00,HenkelPleimling03a,PiconeHenkel04a,CalabreseGambassi05,HenkelPleimling05a,Chamon06a,Hinrichsen06b,LorenzJanke07a,Henkel07a} along similar lines. Here the algebra is relaxed by giving up time-translational invariance, probably reducing the predictive power of the theory even further.

Why is LSI less predictive than conformal invariance in two dimension? In my opinion this issue can be traced back to different symmetry properties. Conformal invariance~\cite{Polyakov70a} generalizes global dilatations combined with \textit{rotations} to a local symmetry. Similarly, LSI generalizes global dilatations combined with \textit{shear transformations} to a local symmetry. However, rotations and shear transformations are very different in character. In the case of conformal invariance, an isotropic equilibrium model is expected to be rotationally invariant \textit{by itself}. Contrarily, as demonstrated in Fig.~\ref{fig:schroedinger}, an anisotropic process is \textit{not} automatically invariant under shear by itself, rather this invariance has to be established manually by adding suitable terms to the generators. At this point the theory of LSI requires an input of extra information which is not needed in conformally invariant systems. The message of this paper is that the ambiguity caused by this additional information reduces the predictive power of LSI. More specifically, it is suggested that this information can be expressed in terms of functions with a single parameter, providing so many degrees of freedom that any two-point function can be reproduced by the theory. 

The present findings make it plausible why LSI was applied successfully to many exactly solvable systems while it continued to fail for certain non-integrable systems. In exactly solvable systems one usually arrives at a partial differential equation which determines the two-point function. This allows one to derive suitable LSI generators in a closed form. For a non-integrable system such as directed percolation~\cite{Hinrichsen06} the scaling function $\Phi$ is a complex object which involves loop corrections to all orders of the underlying field theory. The corresponding LSI generators may exist, but apparently it is impossible to write them down in a closed form. In such cases the attempt to guess a suitable representation and to confirm it numerically by comparing two-point functions is likely to fail.

The results of the present work do not rule out that LSI might have some predictive power on the level of three-point functions. In my opinion this is one of the key issues to be addressed in the future. 

\vspace{3mm}
\noindent
\textbf{Acknowledgement:}\\
I would like to thank M. Henkel for inspiration and interesting discussions.
\newpage

\appendix\section{Consistency of the contruction scheme}

In this appendix it is shown that the generator $X_2$, constrained by Eq.~(\ref{constraints}), determines all generators iteratively in such a way that all commutation relations are satisfied.\\

\noindent
\textbf{A.1: Commutators $[X_n,X_m]=(n-m)X_{n+m}$:}

\def\apply#1#2{\underbrace{#1}_{\mbox{\scriptsize apply~(\ref{#2})}}}

\noindent
To prove these commutation relations by induction, we first notice that 
\begin{eqnarray}
&&\label{AX1}[X_0,X_{-1}]=X_{-1},\hspace{10mm} 
\label{AX2}[X_2,X_{-1}]=3X_1,\\
&&\label{AX3}[X_2,X_0]=2X_2,\hspace{12mm}
\label{AX4}[X_1,X_{-1}]=2X_0\,.\nonumber
\end{eqnarray}
The first relation is always fulfilled, while the second one was used to define $X_1$ in Eq.~(\ref{ConstructX1}). The third and the fourth relation have been used as constraints for $X_2$  in Eq.~(\ref{constraints}) and thus they are satisfied as well. Moreover, the generators $X_n$ have been constructed according to Eq.~(\ref{ConstructX}), hence the commutators
\begin{equation}
[X_n,X_1]=(n-1)X_{n+1} \qquad n=2,3,\ldots
\end{equation}
are valid by construction. Anchored at these relations, the remaining commutation relations $[X_n,X_m]=(n-m)X_{n+m}$ can be proven by induction. To this end let us assume that these commutation relations hold for $n+m=N-1$ and show that they also hold for $n+m=N$:
\begin{eqnarray}
[X_n,X_m] &=& \frac{1}{n-2}[[X_{n-1},X_1],X_m] \nonumber\\
&=& \frac{1}{n-2}\Bigl([\apply{[X_m,X_1]}{ConstructX},X_{n-1}]+[\underbrace{[X_{n-1},X_m]}_{\mbox{\scriptsize induction}},X_1]\Bigr)\\[0mm]
&=& \frac{1}{n-2}\Bigl((m-1)[X_{m+1},X_{n-1}]+(n-m-1)\apply{[X_{n+m-1},X_1]}{ConstructX}\Bigr)\nonumber\\
\label{Nrec}
&=& \frac{1}{n-2}\Bigl((m-1)[X_{m+1},X_{n-1}]+(n-m-1)(n+m-2)X_{n+m}]\Bigr)\nonumber
\end{eqnarray}
This is again a recursion relation with fixed $N$ for an inductive step from $(n-1,m+1)$ to $(n,m)$. More specifically, if the commutation relation $[X_{m+1},X_{n-1}]=(m-n+2)X_{m+n}$ is known to be valid, the above equation implies that the commutator
\begin{eqnarray}
[X_n,X_m] &=& \frac{(1-m)(n-m-2)-(m-n+1)(m+n-2)}{n-2}X_{n+m}\nonumber\\
          &=& (n-m)X_{n+m}
\end{eqnarray}
The same recursion works also in different direction as an inductive step from $(n+1,m-1)$ to $(n,m)$. This twofold recursion scheme allows one to check all commutators of the form $[X_n,X_m]=(n-m)X_{n+m}$ iteratively.\\

\newpage

\noindent
\textbf{A2: Commutators $[X_n,Y_m]=(n/z-m)Y_{n+m}$:}

\noindent
To prove these commutations relations, we first show by induction that
\begin{equation}
\label{XYFirstInd}
[X_n,Y_{-1/z}]=\frac{n+1}{z}Y_{n-1/z}.
\end{equation}
The induction is anchored at $n=-1$ since $[X_{-1},Y_{-1/z}]=[-\partial_t,-\partial_r]=0$. Assuming that the commutation relations are satisfied for $n-1$, i.e.
\begin{equation}
\label{XYrec0}
[X_{n-1},Y_{-1/z}]=\frac{n}{z}Y_{n-1-1/z}
\end{equation}
the same relations holds for $n$ because of
\begin{eqnarray}
[X_n,Y_{-1/z}] &=& \frac{1}{n-2}\,[[X_{n-1},X_1],Y_{-1/z}]\\
&=& \frac{1}{n-2} \Bigl([X_1,\apply{[Y_{-1/z},X_{n-1}]}{XYrec0} +
                        [X_{n-1},\apply{[X_1,Y_{-1/z}]}{ConstructY}]\Bigr)\nonumber \\
&=& \frac{1}{n-2} \Bigl(-\frac{n}{z}\apply{[X_1,Y_{n-1-1/z}]}{ConstructY} +
                        \frac{2}{z}\apply{[X_{n-1},Y_{1-1/z}]}{XYrec0}\Bigr)\nonumber \\
&=& \frac{-\frac{n}{z}\Bigl(\frac{2}{z}-n+1\Bigr)+\frac{2}{z}\Bigl(\frac{n}{z}-1\Bigr)}{n-2}Y_{n-1/z}
\;\;=\; \frac{n+1}{z} \,Y_{n-1/z}\nonumber
\end{eqnarray}
Next, let us assume that the commuation relations
\begin{equation}
\label{XYRec1}
[X_n,Y_{m-1}]=(n/z-m-1)Y_{n+m-1} 
\end{equation}
hold for all $n$ and a given $m$, anchored at $m=-1/z$ by Eq.~(\ref{XYFirstInd}). Then we can prove the remaining relations by a another induction:
\begin{eqnarray}
[X_n,Y_m] &=& \frac{1}{\frac{1}{z}+1-m}\,[X_n,[X_1,Y_{m-1}]]\\
&=& \frac{1}{\frac{1}{z}+1-m}\Bigl([\apply{[Y_{m-1},X_n]}{XYRec1},X_1]+[\apply{[X_n,X_1]}{XXComm},Y_{m-1}]\Bigr)
\nonumber\\
&=& \frac{1}{\frac{1}{z}+1-m}\Bigl(-(\frac{n}{z}-m+1)\apply{[Y_{m+n-1},X_1]}{ConstructY}
                           +(n-1)\apply{[X_{n+1},Y_{m-1}]}{XYRec1}\Bigr)
\nonumber\\
&=& \frac{\Bigl(\frac{n}{z}-m+1\Bigr)\Bigl(\frac{1}{z}-m-n+1\Bigr)+(n-1)\Bigl(\frac{n+1}{z}-m+1\Bigr)}
{\frac{1}{z}+1-m}\,Y_{n+m} \nonumber \\ 
&=& \Bigl(\frac{n}{z}-m\Bigr) \, Y_{n+m} \nonumber
\end{eqnarray}

\newpage

\section{The most important representations}

\def\repsec#1{\vspace{5mm}\noindent{\bf #1}\\[1mm]}

This appendix demonstrates that the most important representations of LSI derived so far can be described within the unified framework of generating kernel functions. 

\repsec{a) Schr{\"o}dinger representation}
%
The standard Schr{\"o}dinger for diffusive systems with $z=2$ is defined by~\cite{Kastrup68a,Niederer72a,Henkel94a}
\begin{eqnarray}
X_n &=& -t^{n+1}\partial_t - \frac{n+1}{2}t^n r \partial_r -\frac{x}{2}(n+1)t^n - \frac{n(n+1)}{4}\mathcal M t^{n-1}r^2 \,,\\
Y_m &=& -t^{m+1/2}\partial_r - (m+1/2)\mathcal M t^{m-1/2} r\,,
\end{eqnarray}
where $\mathcal M$ is the so-called mass parameter. As can be verified easily, this representation is generated by the kernel functions
\begin{equation}
K(t-t',r,r')=\frac12 \mathcal M r^2 \delta(r-r') \delta(t-t')\,,\quad L(t-t',r,r')=0\,.
\end{equation}
%

\repsec{b) Local representation type (i) for arbitrary $z$}
%
This representation is given by~\cite{Henkel02a}
\begin{eqnarray}
X_n &=& -t^{n+1}\partial_t - \frac{n+1}{z}t^n r \partial_r -\frac{x}{z}(n+1)t^n - \frac{n(n+1)}{2}B_{10}t^{n-1}r^z \,,\\
Y_{k-1/z} &=& -t^k\partial_r - \frac{z^2}{2}k B_{10} t^{k-1} r^{z-1}\,
\end{eqnarray}
and reduces to the Schr{\"o}dinger representation for $z=2$. The corresponding kernel reads
\begin{equation}
K(t-t',r,r')=B_{10} r^z \delta(r-r') \delta(t-t')\,,\quad L(t-t',r,r')=0\,.
\end{equation}
%

\repsec{b) Extended local representation type (ii) for $z=2$}
%
The representation~\cite{Henkel02a}
\begin{eqnarray}
X_n &=& -t^{n+1}\partial_t - \frac{n+1}{2}t^n r \partial_r -\frac{x}{2}(n+1)t^n \\&&\nonumber- \frac{n(n+1)}{2}B_{10}t^{n-1}r^2 - \frac{(n^2-1)n}{6}B_{20}t^{n-2}r^4 \,\\
Y_{k-1/z} &=& -t^k\partial_r - 2k B_{10} t^{k-1} r - \frac43k(k-1)B_{20}t^{k-2}r^3\,
\end{eqnarray}
involves both kernels $K$ and $L$:
\begin{eqnarray}
K(t-t',r,r') &=& B_{10} r^2 \delta(r-r') \delta(t-t')\,,\\
L(t-t',r,r') &=& B_{20} r^4 \delta(r-r') \delta(t-t')\,.
\end{eqnarray}

\newpage 

\repsec{c) Local representation type (iii) for $z=1$}
%
The `conformally invariant' representation~\cite{Henkel02a}
\begin{eqnarray}
X_n &=& -t^{n+1}\partial_t - A_{10}^{-1}[(t+A_{10} r)^{n+1}-t^{n+1}] \partial_r \\&&\nonumber-
(n+1)xt^n - \frac{n+1}{2}\,\frac{B_{10}}{A_{10}}[(t+A_{10}r)^n-t^n]\\
Y_{k-1} &=& -(t+A_{10} r)^k\partial_r - \frac{k}{2} B_{10} (t+A_{10} r)^{k-1} r \,
\end{eqnarray}
corresponds to 
\begin{eqnarray}
K(t-t',r,r') &=& \delta(t-t') \Bigl[ A_{10} r^2   \delta'(r-r') + B_{10} r \delta(r-r') \Bigr] \,,\\
L(t-t',r,r') &=& \delta(t-t') \Bigl[ A_{10}^2 r^3 \delta'(r-r') + \frac32 B_{10} A_{10} r^2 \delta(r-r') \Bigr]\,.
\end{eqnarray}
%

\repsec{d) Temporally nonlocal representation for arbitrary $z$}
%
This representation, called type I in Ref.~\cite{HenkelBaumann07b}, is given by
\begin{eqnarray}
X_1 &=& -t^2\partial_t - \frac{2}{z} t r \partial_r -\frac{2x}{z}t \\ \nonumber &&
-(\beta+\gamma)r^2\partial_r^{2-z}-2\gamma(2-z)r\partial_r^{1-z}-\gamma(2-z)(1-z)\partial_r^{-z} \,\\[2mm]
Y_{1-1/z} &=& -t\partial_r - (\beta+\gamma)zr\partial_r^{2-z}-\gamma z (2-z) \partial_r^{1-z}\,.
\end{eqnarray}
It is non-local in space and involves fractional derivatives, 
depending on $z$ even with negative powers. 
Fractional derivatives can always be expressed by integral kernels. As one can see,
this would determine the kernel $K$ while the kernel $L$ vanishes.

\repsec{e) Spatially nonlocal representation for arbitrary $z$}
%
Another recently discovered representation, called type II in Ref.~\cite{HenkelBaumann07b}, is given by
\begin{eqnarray}
X_1 &=& -t^2\partial_t - \frac{2}{z} t r \partial_r -\frac{2x}{z}t - \alpha r^2 \partial_t^{2/z-1} \,\\
Y_{1-1/z} &=& -t\partial_r - \alpha z r \partial_t^{2/z-1}\,.
\end{eqnarray}
As it involves temporal fractional derivatives, this representation is non-local in time.
Again this would determine the form of the kernel $K$ while the kernel $L$ vanishes.\\


\newpage
\noindent{\bf References}

\bibliographystyle{iopart-num}
\bibliography{/home/hinrichsen/Dateien/Literatur/master}

\end{document}